\documentclass[amsmath,amssymb,superscriptaddress,nobalancelastpage,prb,twocolumn]{revtex4}

\usepackage{graphicx}
\usepackage{varioref}
\usepackage{xr-hyper}
\usepackage{xcolor}
\usepackage{hyperref}
\usepackage{ulem}

\begin{document}

\title{A resonant inelastic x-ray scattering study of the spin and charge excitations in the overdoped superconductor La$_{1.77}$Sr$_{0.23}$CuO$_4$}

 \author{C.~Monney}
\affiliation{Swiss Light Source, Paul Scherrer Institut, CH-5232 Villigen PSI, Switzerland}
\affiliation{Physik-Institut, Universit\"{a}t Z\"{u}rich, Winterthurerstrasse 190, CH-8057 Z\"{u}rich, Switzerland}
  \author{T.~Schmitt}
\affiliation{Swiss Light Source, Paul Scherrer Institut, CH-5232 Villigen PSI, Switzerland}
\author{C.~E.~Matt}
\affiliation{Swiss Light Source, Paul Scherrer Institut, CH-5232 Villigen PSI, Switzerland}
\affiliation{Laboratory for Solid State Physics, ETH Z\"{u}rich, CH-8093 Z\"{u}rich, Switzerland}
\author{J.~Mesot }
 \affiliation{Swiss Light Source, Paul Scherrer Institut, CH-5232 Villigen PSI, Switzerland}
\affiliation{Institute for Condensed Matter Physics, \'{E}cole Polytechnique Fed\'{e}rale
de Lausanne (EPFL), CH-1015 Lausanne, Switzerland}
\affiliation{Laboratory for Solid State Physics, ETH Z\"{u}rich, CH-8093 Z\"{u}rich, Switzerland}
  \author{V.~N.~Strocov}
\affiliation{Swiss Light Source, Paul Scherrer Institut, CH-5232 Villigen PSI, Switzerland}
\author{O.~J.~Lipscombe}
\affiliation{H.\ H.\ Wills Physics Laboratory, University of Bristol, Bristol, BS8 1TL, United Kingdom}
\author{S.~M.~Hayden}
\affiliation{H.\ H.\ Wills Physics Laboratory, University of Bristol, Bristol, BS8 1TL, United Kingdom}
 \author{J.~Chang}
 \affiliation{Swiss Light Source, Paul Scherrer Institut, CH-5232 Villigen PSI, Switzerland}
 \affiliation{Institute for Condensed Matter Physics, \'{E}cole Polytechnique Fed\'{e}rale
 	de Lausanne (EPFL), CH-1015 Lausanne, Switzerland}
 \affiliation{Physik-Institut, Universit\"{a}t Z\"{u}rich, Winterthurerstrasse 190, CH-8057 Z\"{u}rich, Switzerland}

\begin{abstract}
We present a resonant inelastic x-ray scattering (RIXS) study of  spin and charge excitations
in overdoped La$_{1.77}$Sr$_{0.23}$CuO$_4$ along two high-symmetry directions.
The line shape of these excitations is analyzed and they are shown to be  highly overdamped. Their spectral weight and damping are found to be 
strongly momentum dependent. 
Qualitative agreement between these observations and a calculated RPA
susceptibility is obtained for this overdoped compound, implying that a significant contribution to the RIXS signal stems from a continuum of charge excitations. 
Furthermore, this suggests that the spin-excitations in the overdoped regime can be captured qualitatively by an itinerant picture.
Our calculations  also predict a new low-energy  spin excitation branch to exist along the nodal direction near the 
zone center. With the energy resolution of the present experiment, this branch is not resolvable but we show that next generation 
of high-resolution spectrometers will be able to test this prediction.
\end{abstract}

\maketitle

\section{Introduction} 
Conventional superconductivity emerges as a result of electron-phonon interaction~\cite{BardeenPR57}.
Information about the phonon excitation spectrum (dispersions and lifetime effects~\cite{KellerPRL2006}) 
are therefore of great importance. 
Similarly, for magnetic superconductors~\cite{ScalapinoRMP12}, there is a strong interest in 
understanding and experimentally revealling the spin excitation spectrum. 
Mapping out the 
detailed evolution of the spin excitation spectrum  across the high-temperature superconducting 
cuprate phase diagram, from
the Mott insulator to the Fermi-liquid ground state, is hence important.
Spin excitations have traditionally been studied by inelastic neutron scattering (INS)~\cite{BirgeneauJPSJ2006,FujitaJPSJ2012}. 
Studies of high-energy  spin excitations~\cite{HaydenPRL96} have, however, 
been challenged by weak neutron cross sections.
 Over the last decade, resonant inelastic x-ray scattering (RIXS) has developed rapidly~\cite{AmentRMP11} and
energy resolution now allows studies of spin excitations~\cite{AmentPRL09,BraicovichPRL09,BraicovichPRL10}. 
RIXS is therefore 
 an attractive complementary technique to neutron scattering. 
 This has, in particular, lead to progress 
in understanding correlated low-dimensional $3d$ and $5d$ electron systems~\cite{KimPRL12,TaconNATP11}.
The spin excitation spectra of insulating 
one- and two-dimensional cuprates have, for example, been studied by soft x-ray RIXS using the 
copper $L_3$-edge~\cite{SchlappaPRL09,SchlappaNAT12,TaconNATP11,GuariseNATC2014,PiazzaPRB2012,GuarisePRL2010,MinolaPRL15}.
In recent years, 
 spin excitations of doped cuprate and pnictide superconductors have also been  investigated~\cite{BraicovichPRL09,BraicovichPRL10,ZhouNATC13,TaconNATP11}.
These studies  suggest that the high energy ($\omega>100$~meV)  spin excitation 
dispersion undergoes little change with doping~\cite{TaconNATP11,DeanNATM13,ZhouNATC13}. This is in strong contrast to the low-energy 
part of the spectrum (studied by INS), that has a strong dependence on impurities~\cite{KimuraPRL03}, magnetic 
field~\cite{ChangPRL2007,ChangPRL09} and doping~\cite{VignolleNATP06,LipscombePRL07}.

We present a systematic RIXS study of the spin and charge excitations
found in overdoped La$_{2-x}$Sr$_{x}$CuO$_4$ (LSCO) $x=0.23$. The line shape of these  excitations 
is analyzed using the response function of a damped harmonic oscillator.  
In this fashion, their dispersion and momentum dependence of spectral weight and damping, $\gamma$, are 
extracted. 
We find that the spectral weight and damping $\gamma$ are displaying a significant momentum dependence. 
The line shape is sharpest around the zone center, whereas the spectral weight increases upon moving towards the zone boundary. 
As reported for Bi-based cuprates~\cite{GuariseNATC2014,DeanPRB14}, we also find a strong nodal / antinodal anisotropy of
spectral weight. 
These observations are captured by susceptibility calculations 
based on the electronic band structure. The model calculation furthermore 
predicts a low-energy spin excitation branch, 
along the ($\pi,\pi$)-direction, which turns out to be particularly pronounced and dispersive in LSCO with $x=0.23$ in comparison to other doped cuprates \cite{GuariseNATC2014,DeanPRL13}. 
Future RIXS experiments with improved energy resolution should test this prediction.

\begin{figure}
	\begin{center}
		\includegraphics[width=0.45\textwidth]{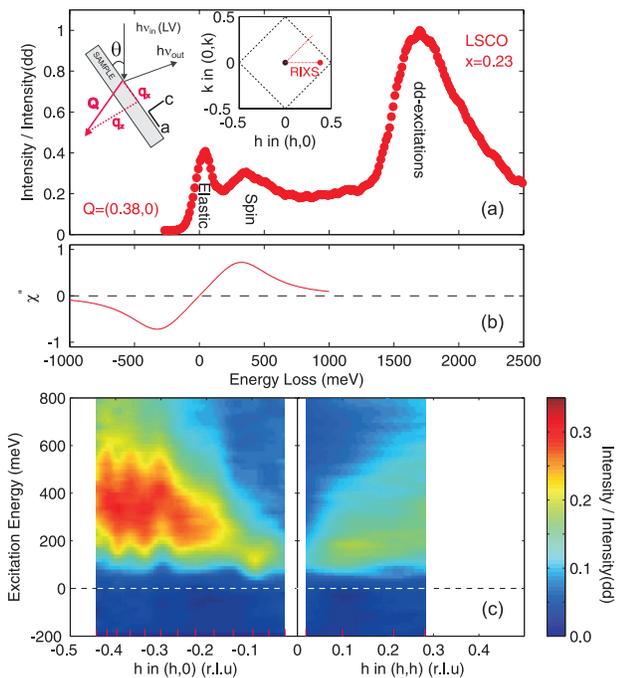}
	\end{center}
	\caption{(a) RIXS spectrum, recorded on overdoped LSCO $x=0.23$ using $\sigma-$polarized light, 
		displays elastic scattering, a low-energy excitation and a $dd$-excitation. The inset shows  
		the scattering geometry and reciprocal space $(h,k)$ schematically. (b) Overdamped response function showing how $\chi^{\prime\prime}\rightarrow 0$ for $\omega\rightarrow0$. 	(c) Interpolated RIXS intensity,
		with elastic scattering subtracted,
		versus momentum $q=(h,0)$, $(h,h)$ and  photon energy loss $\omega$. 
		Red ticks indicate the grid of spectra used for the interpolation.
	}
	\label{fig:fig1}
\end{figure}

 \begin{figure*}
 	\begin{center}
 		\includegraphics[width=0.85\textwidth]{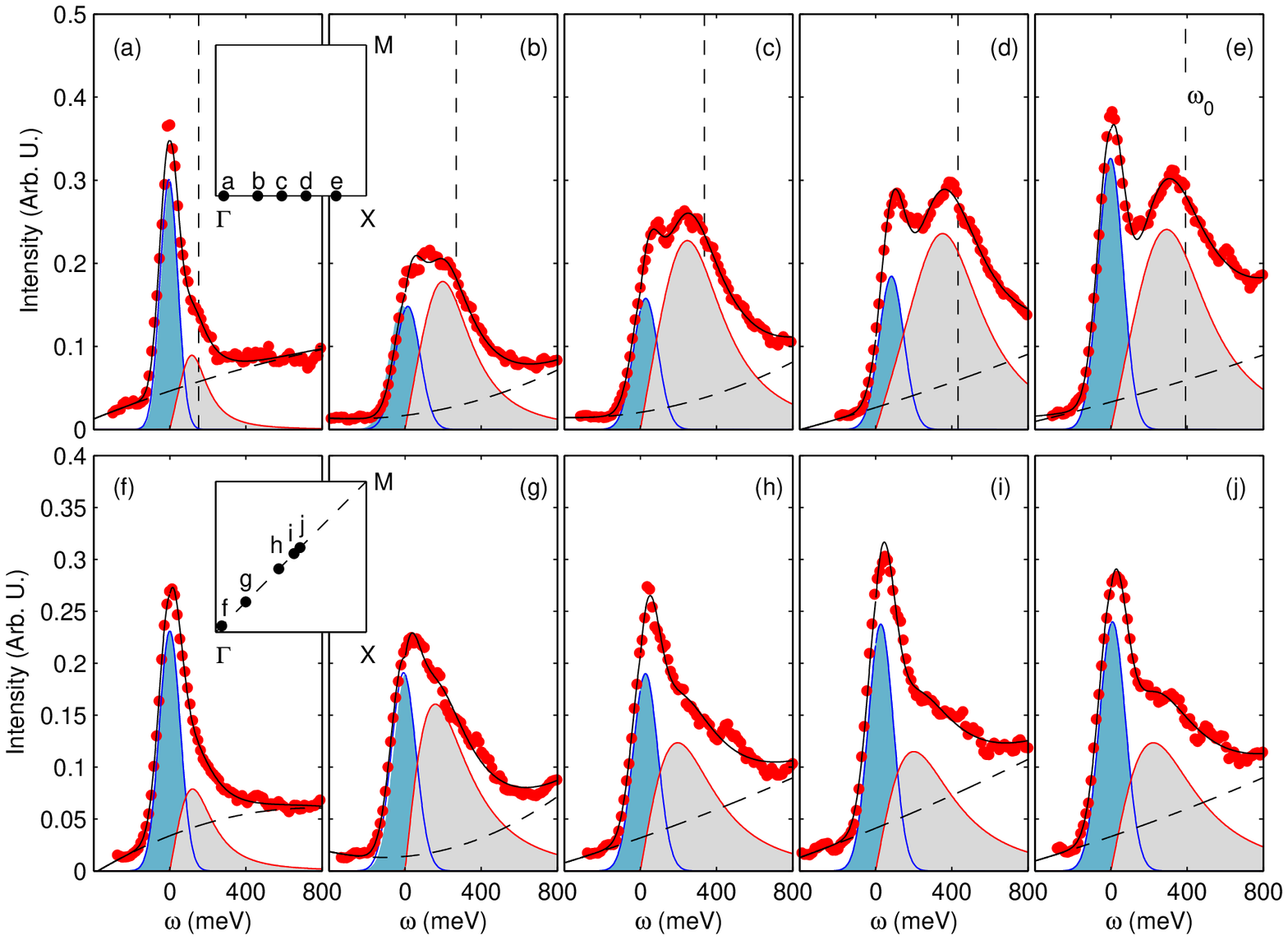}
 	\end{center}
 	\caption{RIXS spectra recorded on LSCO $x=0.23$, at $T= 20$ K, in grazing incidence geometry using $\sigma-$polarized 
 	 light tuned to the Cu $L_3$-edge (930 eV) line. Top panels, (a)-(e), are spectra measured with momenta along $q=(h,0)$ as indicated
 		in the inset of (a).
 		Bottom panels, (f)-(j), displays spectra taken along $q=(h,h)$ as indicated in the inset of (f)-(g). 
 		Blue and gray shaded areas are modelled contributions from elastic and low-energy excitations
 		on top of a cubic background (dashed line). Solid black line is the sum of these contributions.
 		See text for further explanation.}
 \end{figure*}

\section{Method}
 High-resolution RIXS experiments were 
carried out at the ADvanced RESonant Spectroscopy (ADRESS) 
beamline~\cite{ghiringhelliREVSCIINS2006,strocovJSYNRAD2010} at 
the Swiss Light Source (SLS) on
high quality single crystalline  LSCO $x=0.23$ samples~\cite{LipscombePRL07,ChangNATC13,ChangPRB12,FatuzzoPRB2014},
grown by the traveling floating zone method~\cite{KomiyaPRB02}. For the tetragonal crystal 
structure ($a=b\approx3.8$~\AA\ and $c\approx13.2$~\AA), we index the reciprocal space 
by $q=h \bf{a^*}+k \bf{b^*} +\ell \bf{c^*}$ where $\bf{a^*}$ and $\bf{b^*}$ 
point along the Cu-O bonds. Samples were aligned \textit{ex-situ}, using the 
x-ray  Laue technique, in order to 
access 
the scattering planes $(h,0,\ell)$ or $(h,h,\ell)$. 
Cleaving was performed \textit{in-situ} under ultra high 
vacuum conditions ($<5\times 10^{-10}$ mbar)
using a standard top-post technique and the sample was kept at a temperature of 20 K for all measurements. 
At the Cu $L_3$-edge ($\sim 930$ eV), the instrumental energy and momentum half-width-at-half-maximum (HWHM) 
resolutions are $65$ meV and 0.01~\AA$^{-1}$, respectively. The incoming light was $\sigma-$polarized for all measurements.
For each spectrum, the elastic 
line
was obtained by measuring non-resonant elastic scattering from
 polycrystalline carbon containing tape placed just next to the sample~\cite{SchlappaNAT12}.
Reciprocal space positions of the form $(h,0,\ell)$ and $(h,h,\ell)$ were sampled by
changing the grazing incident angle $\theta$, defined in Fig.~1.
The layered cuprates are known to have weak magnetic coupling along the c-axis leading to little dispersion along $\ell$. 
We therefore describe positions using a two-dimensional notation $(h,k)$ to quantify momentum transfer $q$.

\section{Results}
A typical RIXS spectrum  recorded with $\sigma-$polarized light 
at $(h,k)=(0.38,0)$ is shown in Fig.~1a.
As previously reported on the cuprates~\cite{BraicovichPRL10,TaconNATP11,DeanPRL13}, the spectrum consists of 
three features: (1) elastic and quasi-elastic scattering
at $\omega\approx 0$, (2) a low-energy excitation at around 300 meV that has  been interpreted as a spin excitation in the parent compound~\cite{BraicovichPRL09,BraicovichPRL10} and (3) so-called $dd$-excitations at about 1700 meV. 
The $dd$-excitations are in agreement with what has previously been reported on 
LSCO~\cite{GhiringhelliPRL04} and explained by crystal field calculations~\cite{MorettiNJP11,VeenendaalPRL96}. 
Following common practice, all spectral intensities are renormalized to total integrated intensity of these $dd$-excitations, $I_{dd}$ ~\cite{DeanPRL13,DeanNATM13,TaconPRB13}.

As expected,  significant elastic scattering is found near the specular 
condition [$q=(0,0)$] -- see Fig. 2a. The increased elastic scattering near 
the grazing incidence condition $q\approx(0.4,0)$ was previously interpreted as 
a result of a phonon branch~\cite{BraicovichPRL10}. Herein, we make no attempt to 
disentangle contributions from phonons and elastic scattering. 
We also stress that contrary to what was reported~\cite{BraicovichPRL10} in underdoped LSCO $x=0.08$, 
 only one low-energy excitation branch is resolved in our RIXS spectra of overdoped LSCO. Hence, there is 
no evidence for phase separation in our compound.

A systematic compilation of RIXS spectra taken along the $(h,0)$ and $(h,h)$
directions are shown in Fig.~2. For simplicity only the elastic scattering and low-energy excitations 
are shown. In Fig.~1c, the spectral weight originating from 
 these excitations is displayed using a false color scale and 
after subtracting the elastic component.
Without any detailed analysis, following observations can be made. 
(1) Although weaker, their  spectral weight remains finite in 
 the region near the zone center $q=(0,0)$, see Fig.~1c. 
 (2) The spectral weight is weaker
 and the excitations broader and less dispersive along the $(h,h)$ direction.
 A similar dichotomy between ``nodal'' $(h,h)$ and ``antinodal'' $(h,0)$ 
 directions has been reported also for optimally and underdoped 
 Bi$_2$Sr$_2$CaCu$_2$O$_{8+\delta}$~\cite{GuariseNATC2014,DeanPRB14} (Bi2212).
 The less dispersive nodal excitation has also been reported for  overdoped LSCO ($x=0.25$)~\cite{WakimotoPRB2015}.

 \begin{figure}
\begin{center}
\includegraphics[width=0.45\textwidth]{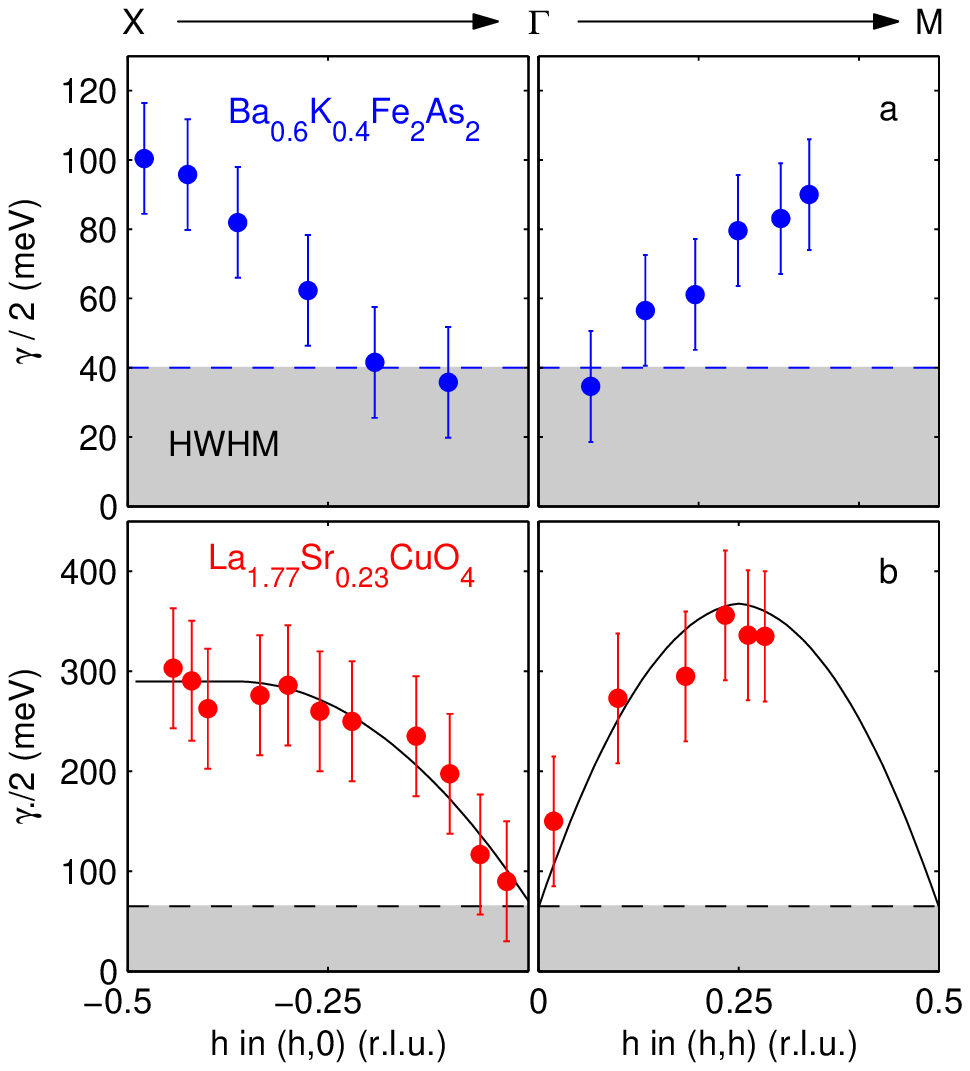}
\end{center}
\caption{Damping $\gamma/2$ (see text) of the low-energy excitations measured by RIXS in LSCO $x=0.23$ 
(this work) and Ba$_{0.6}$K$_{0.4}$Fe$_2$As$_2$~\cite{ZhouNATC13} along the high-symmetry directions $\Gamma$X and $\Gamma$M.
Error bars in bottom panels are set by the applied 
energy resolution (65 meV - HWHM) that is also indicated by a horizontal dash line. 
}

\label{fig:fig3}
\end{figure}

\begin{figure}
	\begin{center}
		\includegraphics[width=0.495\textwidth]{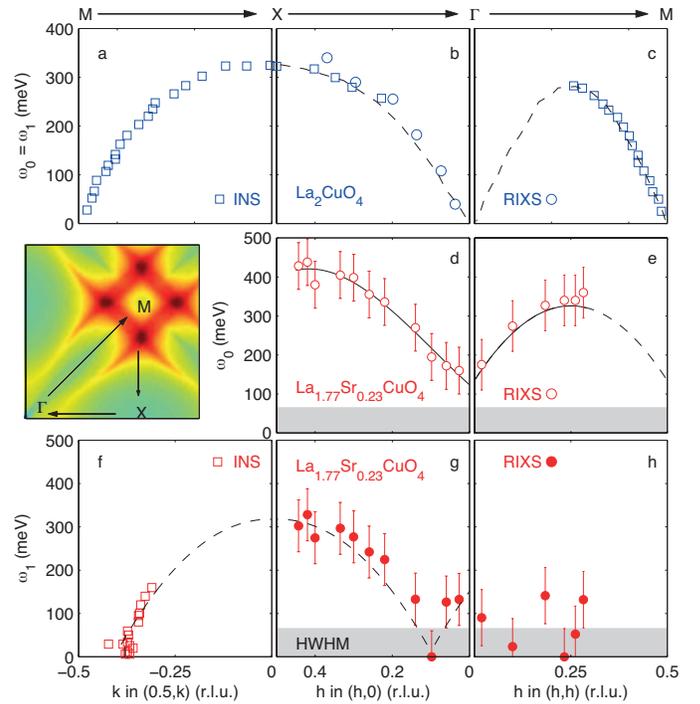}
	\end{center}
	\caption{Comparison of the spin excitation dispersions $\omega_0$ and $\omega_1$ (see text) extracted
		on La$_2$CuO$_4$ (LCO) with the low-energy excitation dispersions on La$_{1.77}$Sr$_{0.23}$CuO$_4$ along high
		symmetry directions as indicated. Data obtained from INS
		and RIXS are displayed by square and circular points respectively. For LCO good agreement between 
		INS [$\square$ - Ref.~\onlinecite{HeadingsPRL10}] and RIXS [$\circ$ - Ref.~\onlinecite{BraicovichPRL10}] is found along the $\Gamma$X-direction.
		No overlap between RIXS (\textbullet~-~this work) and INS ($\square$ - Ref.~\onlinecite{LipscombePRL07}) has been reached for overdoped 
		compositions of LSCO. The inset indicates the high-symmetry directions and displays the 
	calculated static Lindhard susceptibility (for $\omega\rightarrow0$) (see text for further explanation). 
	} 
	\label{fig:fig4}
\end{figure}

 \begin{figure}
\begin{center}
\includegraphics[width=0.45\textwidth]{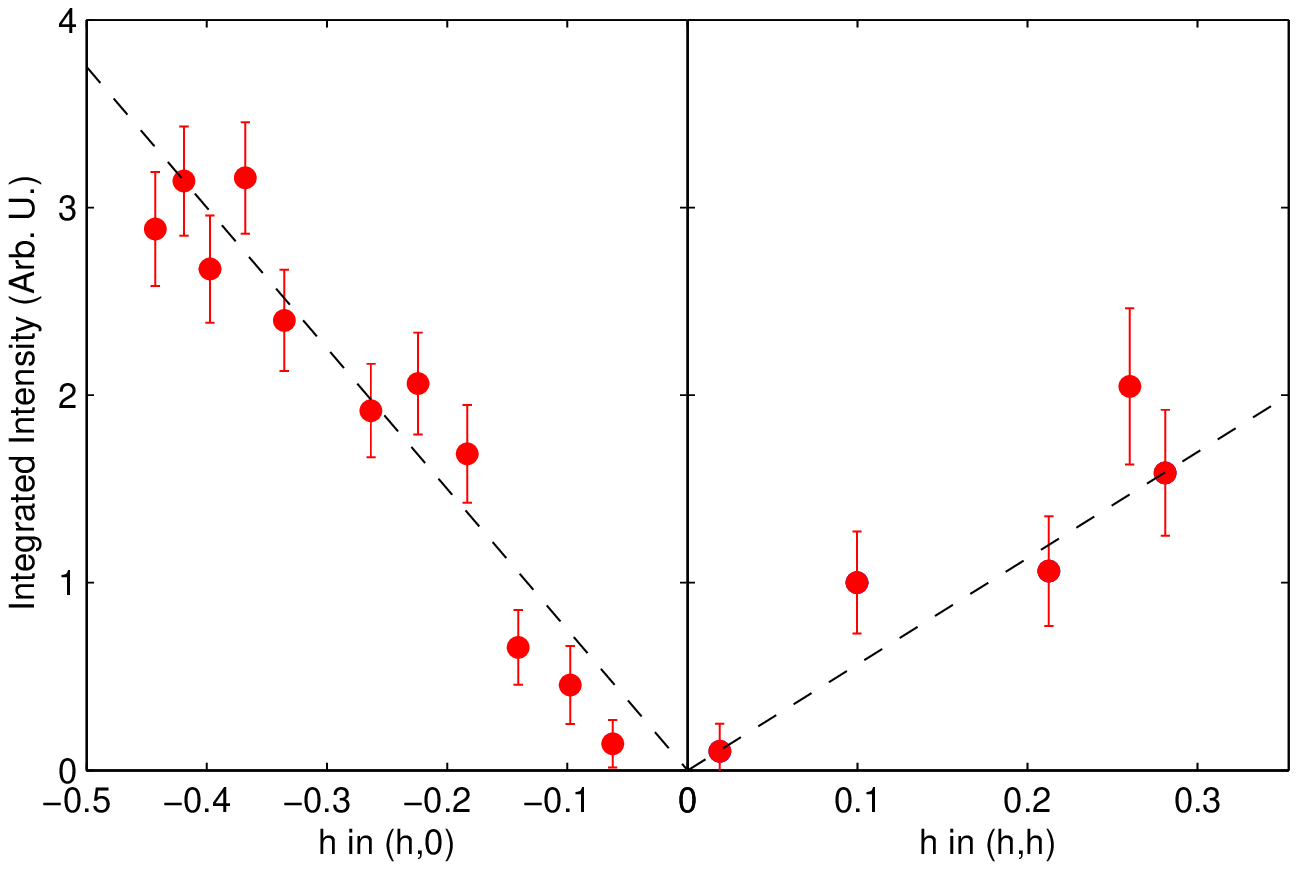}
\end{center}
\caption{Integrated intensity $\chi_0''\cdot\gamma$ of the low-energy excitations measured by RIXS in LSCO $x=0.23$ along the high-symmetry directions $\Gamma$X and $\Gamma$M.
The dash lines are a guide to the eyes.
}

\label{fig:fig4x}
\end{figure}

\section{Analysis}
\subsection{Data modelling}
To model the spectral weight from elastic and low-energy scattering, we use the formula
  $\frac{I}{I_{dd}}(\omega)=G(\omega)+n_B\chi^{\prime\prime}(\omega)$
where $n_B=(1-\exp(\hbar\omega/k_BT))^{-1}$ accounts of the Bose factor. $G(\omega)$ is a Gaussian 
function (to fit the elastic line) on top of a background modeled by a cubic polynomial.
 The response function $\chi^{\prime\prime}(\omega)$ is that of 
a damped hamonic oscillator: 
\begin{align}
\chi^{\prime\prime}(\omega) & =\chi_0^{\prime\prime}  \frac{\gamma\omega }{[\omega^2-\omega_0^2]^2+\omega^2\gamma^2}\nonumber\\
&= \frac{\chi_0^{\prime\prime}}{2\omega_1}\left[\frac{\gamma/2}{(\omega-\omega_1)^2+(\gamma/2)^2}-\frac{\gamma/2}{(\omega+\omega_1)^2+(\gamma/2)^2}\right], \nonumber
 \end{align}
where the damping coefficient $\gamma/2=\sqrt{\omega_0^2-\omega_1^2}$. Considering for a moment only magnetic excitations, this response function spans
two conceptually different regimes. In the limit $\gamma\ll\omega_0$ ($\gamma\rightarrow0$),
$\chi^{\prime\prime} \sim \delta(\omega-\omega_0)-\delta(\omega+\omega_0)$ 
describes coherent propagating magnon excitations with $\omega_0=\omega_1$ as a pole.  
The overdamped limit ($\gamma\approx\omega_0$), in contrast, is characterized 
by  $\chi^{\prime\prime}\propto\omega$ for $\omega\rightarrow 0$ -- see Fig.~1b.
Furthermore, for $\omega_0>\omega_1$, neither of these two energy scales 
reflect the pole of a coherent excitation.
However, as $\chi^{\prime\prime}$ is broadly peaked at $\omega_1$, this energy scale 
is often refered to as the paramagnon excitation energy scale~\cite{TaconNATP11,TaconPRB13,ZhouNATC13}.

Fits to spectra taken at different momenta $q$ along $q=(h,0)$ and $(h,h)$ are shown 
in Fig. 2. Solid lines indicate the elastic (blue) and low-energy excitation (red)
contributions. In this fashion $\gamma$, $\omega_0$, $\omega_1$ and $\chi_0''$ were extracted for LSCO $x=0.23$ 
along the two high-symmetry directions -- see Fig.~3, 4 and 5.
From this analysis, it is found that $\gamma/2$ and $\omega_0$ are comparable for all measured spectra. 
Interpreting the low-energy excitation along $(h,0)$ as a spin excitation, as will be confirmed below, implies that it is overdamped. Along $(h,h)$, the nature of this excitation is less clear and has probably a mixed spin and charge character, which makes the interpretation of its parameters more delicate.
 We note that the damping $\gamma/2$ softens upon moving from the zone boundary towards the zone center (Fig.~3b).
A similar angular dependence has previously been reported in optimally 
doped Ba$_{0.6}$K$_{0.4}$Fe$_2$As$_2$~\cite{ZhouNATC13} (reproduced in Fig.~3a). 
Additionnally, $\omega_0$ disperses upward from the 
zone center and saturates near the zone boundary along both $(h,0)$ and $(h,h)$
directions. A similar dispersion of $\omega_1$ is found along $(h,0)$.
 As $\gamma/2\approx \omega_0$, along the nodal $(h,h)$-direction,
it is difficult to extract $\omega_1$ reliably.
Finally, we observe in Fig.~5 that the integrated intensity $\chi_0''\cdot\gamma$ of the low-energy excitation is weakly anisotropic, as it is larger along the $(h,0)$ direction than the $(h,h)$ direction (for a given absolute value of the momentum $|q|$). 

\subsection{RPA susceptibility calculations}
To analyze the RIXS intensities and neutron scattering spectra, itinerant approaches have been applied~\cite{WeiPRB13,NormanPRB01,EreminPRL05,HePRL11,BenjaminPRL2014,GuariseNATC2014}. These approaches are 
expected to be especially relevant for very overdoped 
cuprates, where the system enters a state with some of the characteristics of a Fermi liquid~\cite{FatuzzoPRB2014,NakamaePRB2003,VignolleNAT2008}.
We have therefore calculated the RPA spin susceptibility $\chi_s(\mathbf{q},\omega)$ for overdoped LSCO, to analyse the low-energy excitations in the paramagnetic state. 
The RPA susceptibility describes the collective magnetic excitations of the itinerant electrons.
Similar to Guarise \textit{et al.}  (Ref.~\onlinecite{GuariseNATC2014}), we obtain here the transverse part of the spin susceptibility as 
\begin{equation}
\chi_s(\mathbf{q},\omega)=\frac{\chi_0(\mathbf{q},\omega)}{1-U\chi_0(\mathbf{q},\omega)},\nonumber
\end{equation}
where $\chi_0(\mathbf{q},\omega)$ represents the Lindhard response function~\cite{EreminPRL05} and $U$ is the local Coulomb interaction. As input to $\chi_0$, we use the single-band tight-binding parametrization~\cite{PavariniPRL01} of the electronic dispersion obtained from ARPES measurements on this 
sample~\cite{ChangNATC13}. The renormalized band width $4t=490$~meV was used and $U$ is chosen to be $1.2t$, so that the susceptibility is not diverging, meaning that the system is far enough from a density-wave instability.

\begin{figure*}
	\begin{center}
		\includegraphics[width=0.85\textwidth]{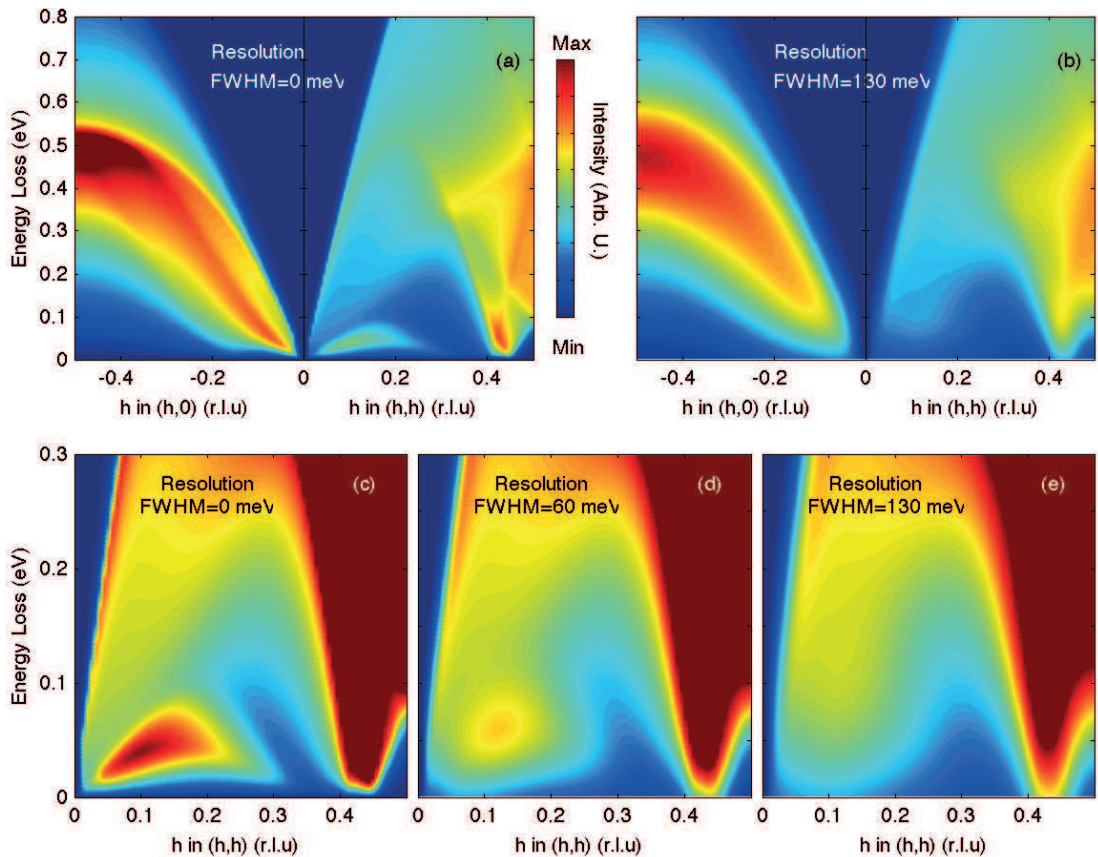}
	\end{center}
	\caption{(a) Calculated RPA susceptibility along the nodal $(h,h)$ and antinodal direction $q=(h,0)$. 
			(b) The  RPA susceptibility convoluted by instrument resolution to make a direct comparison 
			to Fig.~1c. (c-e) A zoom of the low-energy nodal RPA susceptibility. In (d,e) a Gaussian convolution 
			with FWHM = 60 and 130 meV has been applied. This demonstrates that a spectrometer 
			with a 60 meV energy resolution at the Cu $L_3-$edge is sufficient to test the RPA prediction 
			of  low-energy nodal spin-excitations.}
	\label{fig:fig5}
\end{figure*}

In fact, the RPA susceptibility, $\chi_s$, induces moderate modifications of the particle-hole continuum obtained from the Lindhard response function $\chi_0$. 
Along $(\pi,0)$, the dispersion in the particle-hole continuum is renormalized to lower energies (smaller bandwidth) and starts 
to develop a second branch, leading to a
second minimum (softening) at around (0.15,0). In this sense, it can be interpreted as a spin excitation.
Along $(\pi,\pi)$, the main changes occur around the $M$-point, 
where low-energy spin excitations near $(\pi,\pi)$ are reproduced consistently with 
previous susceptibility calculations~\cite{HePRL11} (see Fig. 4, inset). 
Interestingly, a weakly dispersing branch is found in the range $(0,0)\rightarrow (0.2,0.2)$, see Fig.~6a,c.
As it gets stronger with increasing $U$, we interprete it as a spin excitation branch. 
Notice, however, that after convolution of the 
applied instrumental resolution ($\sigma=55$~meV) these detailed features are being smeared out completely (Fig.~6b,e).

\section{Discussion}
The calculated RPA susceptibility contains contributions from both excited particle-hole continuum and 
spin excitations~\cite{BenjaminPRL2014,GuariseNATC2014}. RIXS should be sensitive to both these components.  
The convoluted RPA calculation reproduces 
the most salient observations. First, along the $(\pi,0)$ direction, the 
spectral weight distribution is reproduced quite successfully, compare Fig.~1c with Fig.~6b. Moreover, the 
calculation also produced stronger damping as the excitations disperse towards 
the zone boundary. Second, the susceptibility calculation captures the 
intensity anisotropy between $(0,0)\rightarrow(\pi,0)$ and $(0,0)\rightarrow(\pi,\pi)$. Such a clear anisotropy in the intensity distribution (see Fig.~5) had not been observed previously in other RIXS studies 
on doped cuprates~\cite{GuariseNATC2014,WakimotoPRB2015,DeanPRB14}.
The susceptibility calculation furthermore makes a number of predictions, that can 
be tested by improving the instrumental resolution. Most notable is the 
low-energy excitation branch along $(0,0)\rightarrow(\pi,\pi)$.
Such a low-energy dispersion appeared already in the RPA calculated of other doped cuprates~\cite{GuariseNATC2014,DeanPRB14}, but was not recognized as such, mainly because it was not as distinct as in the present case. We attribute its clear dispersive character here to the specific LSCO electronic structure that has a van Hove singularity in the antinodal region~\cite{ChangNATC13}.
Improving the resolution to have a Gaussian standard deviation $\sigma\sim 25$~meV (FWHM $\sim 60$ meV) would be 
sufficient to resolve this predicted low-energy branch, see Fig.~6d.

In comparison to the case of undoped cuprates\cite{BraicovichPRL10,TaconNATP11}, this analysis shows that the measured excitations in overdoped LSCO are in general broader and their width (see Fig.~3b) has a stronger momentum dependence. This most likely comes from the efficient damping of spin excitations by the electron-hole continuum, as well as from the contribution of electron-hole excitations to the RIXS signal~\cite{BenjaminPRL2014}.

We conclude the discussion by comparing RIXS and INS
studies of LSCO~\cite{BraicovichPRL10,WakimotoPRB2015}.
For the undoped compound, La$_2$CuO$_4$, INS~\cite{HeadingsPRL10}  
and RIXS~\cite{BraicovichPRL10} experiments 
overlap along the $\Gamma$X direction and excellent agreement 
of the measured magnon dispersion is found (see Fig.~4). 
Neutron scattering experiments on doped cuprates are typically restricted 
 -- due to weak cross sections -- to a much narrower range 
 around the (0.5,0,5)-point (indexed $M$)~\cite{LipscombePRL07,WakimotoPRL07}
 where the so-called hour-glass 
 spin excitation dispersion is revealed~\cite{HaydenNAT2004,TranquadaNAT2004,ChristensenPRL2004}. 
The RIXS technique on the other hand has kinematic constrains
 limiting studies to a region  
centered around the $\Gamma$-point. 
For doped cuprates, it is thus difficult to obtain a 
direct overlap of RIXS and INS spectra. 
Within the present RPA calculation the $\Gamma$- and $M$-points are not equivalent.
Caution should therefore be taken 
when comparing neutron scattering data near 
the $M-$point with RIXS data recorded around 
the $\Gamma$-point.

\section{Conclusions}
 In summary, we presented a Cu $L_3-$edge RIXS study of the low-energy spin and charge excitations in overdoped 
La$_{1.77}$Sr$_{0.23}$CuO$_4$. Two high-symmetry directions $(h,0)$ and $(h,h)$ were investigated. 
Spin excitations along $(h,h)$ are strongly damped and the damping 
 is displaying a significant momentum dependence -- larger momentum yields larger damping. Spectral weight also has
 momentum dependence.  Along the antinodal region more spectral weight is 
 found near the zone boundaries and more spectral weight is found in the antinodal direction 
 than the nodal direction. RPA susceptibility calculations starting from
 the experimental observed band structure captures these trends.
This suggests that the measured RIXS signal originates from a mixture of 
 	spin excitations and a continuum of charge excitations.
Furthermore, based on these calculations, we predict a low-energy dispersive spin excitation branch, 
along the ($\pi,\pi$)-direction, which is particularly intense and distinct from other features in the case of La$_{1.77}$Sr$_{0.23}$CuO$_4$. The emerging  ultra-high-resolution 
spectrometers will be able to test this prediction.

\section{Acknowlegdements:} 
C.~M.,  C.~E.~M. and J.~C. acknowledge support by the Swiss National Science Foundation under grant 
number PZ00P$2\_ 154867$, $200021-137783$, PZ00P$2\_142434$, and BSSGI0$\_155873$. 
C. M. also thanks the Alexander von Humboldt Foundation and MaNEP for financial support. S.~M.~H. acknowledges support by the United Kingdom Engineering and Physical Science Research Council under grant number EP$/$J015423$/1$.
  This work was performed at the ADRESS beamline of the SLS at the Paul Scherrer Institut,
  Villigen PSI, Switzerland. We thank the ADRESS beamline staff
   for technical support.

\bibliographystyle{apsrev}
\bibliography{RIXS}

\end{document}